\documentstyle[aps]{revtex}
\begin{document}
\baselineskip 16pt
\parindent0em \parskip1.5ex plus0.5ex minus 0.5ex
\title{PION--PAIR FORMATION AND THE PION DISPERSION\\
RELATION IN A HOT PION GAS}
\author{T.~Alm}
\address{
AG der Max--Planck--Gesellschaft `Theoretische Vielteilchenphysik'
an der Universit\"at Rostock,\\ Universit\"atsplatz 1, D--18051
Rostock, Germany}
\author{G.~Chanfray}
\address{IPN, Universit\'{e} Claude Bernard Lyon I,
43 Bd. du 11 Novembre 1918,\\ F--69622 Villeurbanne Cedex, France}
\author{P.~Schuck}
\address{ISN, IN2P3--CNRS/Universit\'{e} Joseph-Fourier, 53 Avenue
des Martyrs,\\ F--38026 Grenoble Cedex, France}
\author{G.~Welke}
\address{Department of Physics and Astronomy, Wayne State
University, Detroit, Michigan 48202, U.S.A.}
\maketitle
\begin{abstract}
\begin{sloppypar}
The possibility of pion--pair formation in a hot pion
gas, based on the bosonic gap equation, is pointed out and discussed
in detail.  The critical temperature for
condensation of pion pairs (Evans--Rashid transition) is determined
as a function of the pion density. As for fermions, this phase
transition is signaled by the appearance of a pole in the
two--particle propagator.  In bose systems there exists a second,
lower critical temperature, associated with the appearance of the
single--particle condensate. Between the two critical temperatures the
pion dispersion relation changes from the usual quasiparticle
dispersion to a Bogoliubov--like dispersion relation at low momenta.
This generalizes the non-relativistic result for an attractive
bose gas by Evans {\it et al.}  Possible consequences 
for the inclusive pion spectra measured in heavy--ion
collisions at ultra--relativistic energies are discussed.
\end{sloppypar}
\end{abstract}
\hspace{1.8cm}  PACS numbers: 25.75.+r,25.80.Dj,11.10.St
\vfill
\hspace{1.8cm}  Preprint Number: WSU--NP--96--15

\newpage

\section{Introduction}

In heavy ion collisions at very high energies, the matter created
differs qualitatively from what is traditionally studied in nuclear
and elementary particle physics. In the initial stages of the
collision, copious production of gluons and quarks in a large volume
leads to a rapid increase in the entropy, and the distinct possibility
of a new phase of matter characterized by deconfined degrees of
freedom. One therefore hopes that relativistic heavy ion experiments
can provide insight into the structure of the QCD vacuum,
deconfinement, and chiral symmetry restoration.

The hot transient system eventually evolves into a gas of hadrons at
high energy densities, whose properties may be studied theoretically
using, for example, hadronic cascades \cite{RQMD,ARC,WB}. In
principle, these models provide information on the early, dense phase
by tracing the evolution of the system from hadronization to
freeze--out.  Of course, in ultrarelativistic heavy ion collisions,
most the of the produced secondaries are pions. For example, in
central Au+Au collisions at center--of--mass energies of $200~{\rm
  A~GeV}$ estimates from the FRITIOF event generator suggest that
$\sim 4000$ pions per isospin state might be produced.  Further,
recent measurements \cite{zajc2} at lower energies and comparison to
simulations \cite{RQMD2} show that freeze--out source sizes probably
deviate quite drastically from a simple multiplicity scaling law:
present calculations indicate $10$--$20~{\rm fm}$ Au+Au source radii at
$\sqrt{s}=200~{\rm A\cdot GeV}$.  In any event, these high energy
collisions might well create highly degenerate bose systems, and even
possibly Bose--Einstein condensates (BEC).  Since practical
conclusions from dynamical simulations \cite{BDSW} depend
qualitatively on the effect of the medium on particle interactions
\cite{ACSW,BBDS}, one needs to better understand the properties of
such degenerate systems of pions within the environment of a
relativistic heavy ion collision.

Non--relativistically, the problem of interacting, degenerate
bose systems has been discussed extensively by several authors.
Evans and Imry \cite{Evans1} established the pairing theory of a
bose superfluid in analogy to the BCS theory of superconductivity.
For an attractive interaction, the resulting gap equation may have a
non--trivial solution. Further, though, there appears the possibility
of having a macroscopic occupation of the $k=0$ particle state when
the corresponding BCS quasiparticle energy vanishes. In turn, this
leads to a spectrum which is linear and gapless in the long
wavelength limit \cite{Evans1}.  In a second paper, Evans
and Rashid \cite{Evans2} rederived the equations of
Ref.~\cite{Evans1} using the  Hartree--Fock--Gorkov decoupling
method, and solved them for the case of superfluid helium.
This boson pairing theory has been generalized by
D\"orre {\it et al.} \cite{Haug}, who carried out a thermodynamic
variational calculation with a trial Hamiltonian containing a
c--number part. An extensive discussion on the boson pairing problem
is also given by Nozi{\`e}res and Saint James \cite{StJNoz}.

It has further been shown by Stoof \cite{Stoof1} and, independently,
Chanfray {\em et al.} \cite{CSW} that the critical temperature $T_{\rm
  c}$ for the transition from the normal phase to the phase with a
non--vanishing gap (the Evans--Rashid transition) is given by a
``Thouless criterion'' \cite{Bogoliubov,Thouless} for the bosonic
$T$--matrix in the quasiparticle approximation, in analogy to the
fermion case.  Moreover, it has been demonstrated that there exists a
second critical temperature $T_{\rm BEC}<T_{\rm c}$, where the
condition for the macroscopic occupation of the zero momentum mode of
Ref.~\cite{Evans1} is fulfilled \cite{Stoof1,CSW}. 
The mechanism is the same as for
Bose--Einstein condensation in the ideal bose gas \cite{Stoof1}.

Here we wish to consider $\pi$--$\pi$ interactions in the presence of
a dense and hot pion gas along the lines of a previous approach
\cite{ACSW,CSW}. 
We address the question of pion pair formation and
the pion dispersion relation in a thermal medium, first in a qualitative way
(section II), then in a more detailed numerical calculation with a
realistic two pion interaction (section III). As we shall
see in section IV, the in--medium $2\pi$ propagator exhibits a pole above a
certain critical temperature, signaling a possible instability
with respect to pion pair formation. 
We conclude in section V with a discussion of
the effect in high energy heavy ion collisions.

The effects we present here require rather large phase space densities
for the pions, but are independent of whether full thermal
equilibration has been reached. Nonetheless, we choose to couch the
discussion in terms of thermal language, because it is convenient, but
also because the actual situation is probably not too far removed from
it. Dynamical calculations \cite{WB,BDSW} show that a high degree of
thermal equilibration is quite reasonable.
Chemical equilibration, on the other hand, may well cease 
at later stages of the system's
evolution and lead to a condensation of pions 
in the adiabatic limit. Of course, the system actually expands
rather rapidly, but nonetheless large chemical potentials 
($\mu \sim 130~{\rm MeV}$) may be built up by freezeout ($T \sim
100~{\rm MeV}$). One might thus expect large phase space occupation 
numbers at low momenta, which drive the pion pair formation that we
discuss here.

\section{The Evans--Rashid transition in a hot pion gas}

In order to treat the gas of interacting pions we will use the
boson pairing theory of Evans {\it et al.} \cite{Evans1,Evans2}.  
In analogy to the fermion (BCS) case, they obtain a system of 
coupled equations for the gap energy and the density by linearizing
the equations of motion.  The
usual Thouless criterion for fermions can be established
analogously for the bose system, and yields the critical temperature
below which the gap equation begins to exhibit non--trivial solutions.
However, in contrast to the fermion case, a second, lower, critical
temperature appears at which the quasiparticle energy vanishes at zero
momentum. This temperature is associated with the Bose--Einstein
condensation (BEC) of single bosons, in analogy to the ideal bose gas,
as discussed in Ref.~\cite{CSW}, and in detail for atomic cesium
in Ref.~\cite{Stoof1}. An interesting feature of the
formalism developed by Evans {\it et al.} \cite{Evans1} is that below
the second critical temperature the dispersion relation for the single
bosons is of the Bogoliubov form, {\it i.e.}, linear, or phonon--like,
for small momenta.

In this section, we illustrate these remarks
concerning the Evans--Rashid transition for a pion
gas in a qualitative way, returning to a more detailed numerical
calculation in section~III. While relativistic kinematics is taken
into account, corrections from backward diagrams are ignored. We shall
see in section~III that such an approximation is justified for the
physical regions in which a solution to the gap equation exists. For
clarity in this preliminary discussion, we shall also
generally neglect the $k$-dependence of the self--energy,
$\Sigma(k)$, and further condense it into the chemical potential.

The gap equation for the pion pairs is derived in the appendix,
using Gorkov decoupling:
\begin{equation}\label{pgap}
\Delta(k)\;=\; -{1\over 2}
\,\sum_{{\vec k}^\prime}\: V(k,k^\prime,E=2 \mu)\,
\frac{\Delta(k^\prime)}{2 E(k^\prime)} \,\coth{\frac{E(k^\prime)}{2T}}~ ~,
\end{equation}
where the quasiparticle energy is given by
\begin{eqnarray}\label{Ek}
E(k) \;=\; \sqrt{\epsilon(k)^2-|\Delta(k)|^2}~ ~ ~,
\end{eqnarray}
with $\epsilon(k) \equiv \omega(k)-\mu$, and where
$\omega(k)$ is the free pion dispersion.  The $\coth$--factor
represents the medium effect for a thermalized pion gas at temperature
$T$ and chemical potential $\mu$, and $V(k,k^\prime,E)$ is the as yet
unspecified bare two--particle interaction.  The corresponding pion
density is
\begin{equation}\label{pdens}
n\;=\;\sum_{{\vec k}^\prime} \: \bigg [ \frac{\epsilon(k^\prime)}
{2 E(k^\prime)}\,\coth{\frac{E(k^\prime)}{2T}} \:-\: \frac{1}{2} 
\bigg ]~ ~.
\end{equation}
In spite of the formal similarities of Eq.~(\ref{pgap}) with the
corresponding fermionic gap equation, there are important differences:
For bosons, the $\Delta^2$ is subtracted in $E(k)$ (fermions:
added), and the temperature factor is a hyperbolic cotangent (fermions: tanh).

We discuss the solution to the gap equation for decreasing
temperature, at a fixed value of the chemical potential $\mu$.  The
possibility of a finite chemical potential in a pion gas has been
pointed out in the introduction. At very high temperatures, the gap
equation (\ref{pgap}) has only the trivial solution $\Delta\!=\!0$,
and Eq.~(\ref{pdens}) is the usual quasiparticle density. The
dispersion relation is also of the usual form
\begin{equation}\label{df}
\lim_{\Delta\rightarrow 0} E(k) = \epsilon(k)
=\omega(k)-\mu=\sqrt{k^2+m^2_{\pi}}-\mu~ ~.
\end{equation}
With decreasing temperature, however, a critical temperature $T_c^u$ is
reached, at which the gap equation (\ref{pgap}) first exhibits a
non--trivial solution $\Delta \neq 0$.  The value of $T_c^u$ may be
found by linearizing the gap equation, {\it i.e.}, setting $E(k)
\approx \epsilon(k)$ in Eq.~(\ref{pgap}). We return to this point in
section~III, showing that the resulting equation for $T_c^u$
is identical to the condition for a pole in the two--pion $T$--matrix
at the particular energy $E\!=\!2\mu$ (for total momentum
$\vec{K}\!=\!0$). Thus we have a bosonic version of the well--known
Thouless criterion for the onset of superfluidity in fermion
systems with attractive interactions.

Below the critical temperature $T_c^u$ the order parameter $\Delta$
becomes finite, and the corresponding dispersion relation is now
given by Eq.~(\ref{Ek}).
As the temperature drops further, $|\Delta|$ increases to a point
where the condition $|\Delta(k=0)|=|m_{\pi}-\mu|$ is reached.  This is
the maximum possible value of $|\Delta|$, since otherwise imaginary
quasiparticle energies result. It defines a second critical
temperature $T^\ell_c$, below which the occupation $n_0$ of the zero
momentum state becomes macroscopically large because
$E(k)\!\rightarrow\! 0$ for $k\!\rightarrow\! 0$ \cite{Evans1}.  The
possibility of a macroscopic occupation of the $k=0$ mode below
$T^\ell_c$ follows from the pion density Eq.~(\ref{pdens}): for
$E(k=0)\!=\!0$, the $k\!=\!0$ contribution to the density must be treated
separately, as in the case of the ideal bose gas.  A similar comment
applies to Eq.~(\ref{pgap}) for the gap, so that we obtain the 
two inhomogeneous equations
\begin{eqnarray}
\Delta(k)&=& -\frac 12 \,V(k,0,2 \mu)\:n_0
\:-\:\frac 12\,\sum_{{\vec k}^\prime \neq 0}\:V(k,k^\prime,2 \mu)\,\frac
{\Delta(k^\prime)}{2 E(k^\prime)} \coth{\frac{E(k^\prime)}{2T}}~ ~,
\label{ggu} \\
n&=&n_0\:+\:\sum_{{\vec k}^\prime \neq 0}\:\bigg [
\frac{\epsilon(k')} {2 E(k')}\coth{\frac{E(k')}{2T}}\;-\;
\frac{1}{2}\bigg ]~ ~ ~. \label{dgu}
\end{eqnarray}
In contrast to the ideal bose case, the condensation of quasiparticles
happens at $\mu\!<\!m_\pi$, because of the finite value of the gap.
Below $T_c^\ell$ the dispersion relation is given by
\begin{eqnarray}\label{duu}
E(k)&=&\sqrt{\omega(k)^2-2 \omega(k) \mu+2
\mu m_{\pi}-m^2_{\pi}}~ ~, \nonumber \\
&\approx &\sqrt{2(m_\pi-\mu)\frac{k^2}{2m_{\pi}}\:+\:
\frac{\mu}{m_\pi} (\frac{k^2}{2m_{\pi}})^2}~ ~,\label{dgur}
\end{eqnarray}
in the small $k$, non--relativistic approximation.
Thus, instead of the usual $k^2$--behavior, the pion dispersion is
linear in the long wavelength limit.
Eq.~(\ref{dgur}) may be rewritten in the more usual form of the
well--known Bogoliubov dispersion relation
\cite{Bogoliubov} for a weakly interacting bose gas:
\begin{equation}\label{Bdr}
E(k)\;=\; \sqrt{|V(0,0,2 \mu)|\, n_0\:
\frac{k^2}{2m_{\pi}}   \;+\; O(k^4)}~ ~ ~.
\end{equation}
Here, we have used $m_{\pi}-\mu= -V(0,0,2 \mu)\, n_0/2$, which follows from
Eq.~(\ref{ggu}) for sufficiently low temperatures.


\section{Numerical results for the gap equation}

We now consider the qualitative discussion of the previous section in
more detail, by numerically solving our system of equations for a
realistic pion--pion interaction in the $\ell=I=0$ channel. We choose
a rank--2 separable $\pi$--$\pi$ interaction inspired by the linear
$\sigma$--model (see appendix) which possesses all the desired low
energy chiral properties, as is explicitly discussed in
Ref.~\cite{Drop}.  For vanishing incoming total momentum, ${\vec
  K}=0$, it reads (see Eq.~(\ref{InvMB}))

\begin{eqnarray}
\langle {\vec k}, -{\vec k} \mid V_{I=0}(E) \mid {\vec k}^\prime,
-{\vec k}^\prime
\rangle &=& \frac {v({\vec k})}{2\omega(k)}\: \frac
{M_\sigma^2-m_\pi^2} {f_\pi^2}\: \bigg [ 3\,\frac
{E^2-m_\pi^2}{E^2-M_\sigma^2} \:+\: \frac
{4\omega(k)\omega(k^\prime)
  - 2m_\pi^2}{M_\sigma^2} \bigg ] \; \frac {v({\vec k}^\prime)}
{2\omega(k^\prime)}~ ~ ~,\nonumber \\
&=& \frac {1}{2\omega(k)} \; \langle k \mid {\cal M}_B(E) \mid
k^\prime \rangle \; \frac {1}{2\omega(k^\prime)}~ ~ ~, \label{0PV}
\end{eqnarray}
where, for later convenience, we have introduced the bare invariant
matrix
\begin{equation}
\langle k \mid {\cal M}_B(E) \mid k^\prime \rangle \;\equiv\;
\lambda_1(E)\:v_1(k)v_1(k^\prime) \;+\; 
\lambda_2\:v_2(k)v_2(k^\prime) ~ ~ ~\label{1PV}
\end{equation}
with notation  $v_1(k)\equiv v(k) = [1+(k/8m_\pi)^2]^{-1}$,
$v_2(k) = (\omega(k)/m_\pi) v(k)$, and
\begin{equation}
\lambda_1(E) \equiv \frac {M_\sigma^2-m_\pi^2}{f_\pi^2} 
\; \bigg [ 3 \: \frac {E^2-m_\pi^2}{E^2-M_\sigma^2} \:-\: \frac
{2m_\pi^2}{M_\sigma^2} \bigg ]~ ~, ~ ~ ~ ~ 
\lambda_2 \equiv \frac {M_\sigma^2-m_\pi^2}{f_\pi^2}
\; \frac{4m_\pi^2}{M_\sigma^2}  ~ ~ ~. \label{Vab}
\end{equation}
The form factor $v(k)$ and $\sigma$--mass $M_\sigma=1~{\rm GeV}$ are
fit to experimental phase shifts, as in Ref.~\cite{Drop}.  For free
$\pi^+$--$\pi^-$ scattering this force yields, when used in the
$T$--matrix (see below), a scattering length which vanishes in the
chiral limit, as it should. This feature induces off--shell repulsion
below the $2\pi$--threshold in spite of the fact that the positive
$\delta^0_0$ phase shifts indicate attraction. It is remarkable that
the gap equation still shows a non--trivial solution, signaling pion
pair formation, as we will show later. It is evident that bound pair
formation, or even larger clusters of pions, can deeply influence the
dynamics of the pion gas.

In the sigma channel $(\ell=0,I=0)$
we rewrite the gap equation (\ref{pgap}) as
\begin{equation}\label{gaps1}
\Delta(k) \;=\; -\frac 12 \:\int \! \frac {d^3k^\prime}{(2\pi)^3}
\: \langle {\vec k}, -{\vec k} \mid V_{I=0}(E\!=\!2\mu) \mid
{\vec k}^\prime, -{\vec k}^\prime
\rangle \; \frac {\Delta(k^\prime)} {2E(k^\prime)}
\; \coth \frac {E(k^\prime)}{2T}~ ~ ~,\label{gapeqn}
\end{equation}
With the form of our interaction solutions of this equation
may be written as
\begin{equation}
\Delta(k) \;=\; \frac {m_\pi}{\omega(k)} \: \bigg [  \delta_1
\,v_1(k) \:+\: \delta_2\, v_2(k) \bigg ]~ ~ ~,\label{SolnForm}
\end{equation}
and Eq.~(\ref{gapeqn}) reduces to two coupled non--linear equations
for the ``gap strengths'' $\delta_1$ and $\delta_2$.  For a
non--trivial solution, one can show that $\delta_1 > - \delta_2 > 0$.
We also note that while $\lambda_2$ is always repulsive,
$\lambda_1(E)$ is attractive at $E\!=\!2\mu$ only if $\mid\! \mu\!
\mid\, >M_\sigma m_\pi/2\sqrt{3M_\sigma^2-2m_\pi^2} \sim 40~{\rm
  MeV}$.  This inequality is also the formal condition for a solution
to the gap equation to exist at some temperature. Intuitively, we
require at least some attraction because, as we shall see, a solution
to the gap equation is connected to the existence of a pole in the
$T$--matrix.  We note that the repulsive part of the $\pi$--$\pi$
interaction Eq.~(\ref{0PV}) helps to avoid collapse. This is
different from our previous calculation \cite{CSW}, which was
performed with an entirely attractive interaction. The presence of
this repulsion is a consequence of chiral symmetry and PCAC
\cite{Drop}.

In the previous section, we introduced the critical temperatures 
$T_c^u$, at which the gap vanishes, and $T_c^\ell$, where the gap has
reached its maximum value and quasiparticle condensation occurs.
Fig.~1 shows the numerical results for these temperatures
in the $\mu$--$T$ plane.
The $T_c^u$ (solid line) are obtained by linearizing Eq.~(\ref{gapeqn}):
\begin{equation}
\Delta(k) \;=\; -\frac 12 \:\int \! \frac {d^3k^\prime}{(2\pi)^3}
\: \langle {\vec k}, -{\vec k} \mid V_{I=0}(E\!=\!2\mu) \mid
{\vec k}^\prime, -{\vec k}^\prime
\rangle \; \frac {\Delta(k^\prime)} {2\epsilon(k^\prime)}
\; \coth \frac {\epsilon(k^\prime)}{2T_c^u}~ ~ ~,\label{thou1}
\end{equation}
while the $T_c^\ell$ (dashes) result when the gap
strength increases to a point where $E(k=0)=0$,
{\it i.e.}, $m_\pi-\mu = \delta_1+\delta_2$.
At high temperatures $T>T_c^u$ (region III), 
the system is in the normal state with no gap, while below
the dashed line, $T<T_c^\ell$ (region I), 
there is macroscopic occupation of the
$k=0$ mode. For $T_c^\ell<T<T_c^u$ (region II), 
non--trivial gap solutions exist. Notice
that for physically realistic solutions ($T<200~{\rm MeV}$,
say) we have $\mu\; {\buildrel < \over \sim} \; m_\pi$, and $\omega-m_\pi
\ll m_\pi$, and, in hindsight, are justified in neglecting 
relativistic corrections to the gap equation (see appendix).

Fig.~2 shows the gap strengths $\delta_1$ (solid line) and $-\delta_2$
(dashes) versus temperature for a fixed chemical potential
$\mu=135~{\rm MeV}$. Again, we see that at high
temperatures, in region~III, only the trivial solution
$\delta_1=\delta_2=0$ exists. As the temperature drops to
$T_c^u\sim 123~{\rm MeV}$, the order parameter $\Delta$ switches on,
and we have a transition to a paired state in region II (see
discussion below).  Finally, at $T=T_c^\ell \sim 77~{\rm
  MeV}$, the gap has reached its maximum value $\delta_1+\delta_2 = m_\pi
-\mu \sim 3~{\rm MeV}$ and quasiparticles condense in the
lowest energy mode in region I.

The change in the pion dispersion relation $E(k)$ is investigated 
in Fig.~3 in the temperature range $T_c^\ell\le T\le T_c^u$, for a fixed
chemical potential of $\mu=135~{\rm MeV}$. At $T=T_c^u\sim 123~{\rm
  MeV}$ (solid line), and above, 
we simply have the normal--state pion dispersion
relation $\epsilon(k)= \omega(k)-\mu$.  With decreasing temperature the
influence of the finite gap becomes visible at long wavelengths: The
dot--dashed line shows $E(k)$ for $T=115~{\rm MeV}$. A further drop in the
temperature to $T=T_c^\ell\sim 77~{\rm MeV}$ qualitatively changes
the character of the pion dispersion relation 
to a linear, phonon--like dispersion at small $k$.

\section{The in--medium $\pi\pi$ scattering matrix}

We turn now to a discussion of the $T$--matrix ${\cal
 M}_{I=0}(E,K)$ for a pion pair with total momentum $K = \mid\!{\vec
 K}\!\mid$ with respect to a thermal medium.  Writing the
on--shell $T$--matrix ({\it c.f.} Eq.~(\ref{A:MSol})) as
\begin{equation}
\langle k^* \mid {\cal M}_{I=0}(E,K) \mid k^* \rangle \;=\; 
\sum_{i=1}^2 \: \lambda_i(s) \: v_i(k^*)\,\tau_i(k^*; s,K)~ ~ ~, 
\label{MSol}
\end{equation}
where $k^{*\,2}=s/4-m_\pi^2$ and $s=E^2-{\vec K}^2$ is the square of
the total c.m. energy, the Lippmann--Schwinger equation
(\ref{A:Tmat2}) becomes a set of two linear equations for the functions
$\tau_i$: 
\begin{equation}
\sum_{j=1}^2 \; \bigg [ \delta_{ij}\:-\; \lambda_j(s)\,g_{ij}(s,K)
\bigg ] \:\tau_j(k^*; s,K) \;=\; v_i(k^*)~ ~,~ ~ ~ ~i=1,2 \label{MME}
\end{equation}
with
\begin{equation}
g_{ij}(s,K) \;\equiv\; \frac 12 \: \int \frac{d^3k}{(2\pi)^3} \;
v_i(k) \: \frac {1}{\omega(k)} \, \frac {\langle 1+f_++f_- \rangle}
{s-4\omega_k^2 +i\eta} \: v_j(k)~ ~ ~.\label{Deffij}
\end{equation} 
Here, $\langle 1+f_++f_- \rangle$ denotes an average over the angles
of the c.m. relative momentum of the pair. For thermal occupation
numbers it is given by Eq.~(\ref{A:Angle}), and reduces to unity in
free space and $\coth(\omega(k)-\mu)/2T$ for vanishing total 
momentum ${\vec K}$. We note that Eq.~(\ref{MME}) does not
incorporate the non--linear effect of the gap.\label{Concern1}

The solid line in Fig.~4 shows $\mid\!{\cal M}_{I=0}\!\mid^2$ for free
space scattering. Compared to our previous calculation
\cite{ACSW}, the $T$--matrix is relatively flat above the resonance,
this being due to the repulsion in the interaction at high energies. 
The short dashes give $\mid\!{\cal M}_{I=0}\!\mid^2$ in a thermal 
bath of $T=100~{\rm MeV}$ and $\mu=135~{\rm MeV}$, for $K=0$. The 
medium strongly suppresses the cross
section, an effect that also occurs in the $(I=1,\ell=1)$ $\rho$--channel
\cite{ACSW,BBDS}. At high c.m. energies, the phase space occupation
becomes negligible, and the cross section returns to its free space
value. The three remaining curves show results in the same thermal
bath, but for
$K=200~{\rm MeV}/c$ (long dashes), $1~{\rm GeV}/c$ (dot--dashed),
and $3~{\rm GeV}/c$ (dotted). As $K$ increases, the pair is boosted
more and more out of the occupied phase space of the medium, and the
cross section again returns to its free space value. We also see a
threshold behavior in Fig.~4: as $K$ becomes larger, a resonance
peak emerges from below the threshold which continues to shift up in
energy and strengthen until it coincides with the free scattering
peak. We shall see below that this is the continuation of an upward 
shift of the Cooper pole in the $T$--matrix with decreasing phase space
occupation \cite{CSW}.\label{Concern2}

We consider now the existence of poles in the $T$--matrix, first
for the special case of zero total momentum, $K=0$ \cite{CSW}, and define
the determinant function
\begin{equation}
F_{\mu,T}(E) \;\equiv\; -\,
{\rm det} \:\bigg [ \delta_{ij}\:-\; \lambda_j(E)\,g_{ij}(E)
\bigg ]~ ~ ~. \label{DetFunc}
\end{equation}
This function is shown in Fig.~5 for five different temperatures (solid
lines) at a fixed pion chemical
potential of $135~{\rm MeV}$. The intersection of these curves with zero
(horizontal dashes) below $2m_\pi$ (the bound state
domain) gives the pole position. We see that a pole always occurs 
provided  the temperature lies above some critical
value $T_0^\ell \approx 47~{\rm MeV}$, for which the pole is at
threshold and ceases to exist.
This  $T_0^\ell$ is close to 
the lower critical temperature for the gap, $T_c^\ell$,
where the excitation spectrum vanishes at $k\!=\!0$ and quasiparticles
begin to condense as singles. Thus, the
bound state and gap solution disappear at a similar critical
temperature; differences are ascribable to the fact that we use free
quasiparticle energies in the $T$--matrix.\label{point1}

There is a second special temperature $T_0^u$, for which
a pole exists at $E=2\mu$ (see Fig.~5). It is identical
to the upper critical temperature $T_c^u$ at which the gap vanishes, as
may easily be seen by rewriting the $T$--matrix for $E$ near $2\mu$,
\begin{equation}
\langle {\vec k}, -{\vec k} \mid T_{I=0}(E) \mid {\vec k}^\prime, -{\vec
  k}^\prime \rangle \;\equiv \; Z(k)\: \frac {1}{E-2\mu} \:
Z(k^\prime)~ ~ ~.\label{TnearPole}
\end{equation}
In the non--relativistic limit, $Z(k)$ follows as (see appendix)
\begin{eqnarray}
Z(k) &=& -\frac 12 \:\int \! \frac {d^3k^\prime}{(2\pi)^3} \: \langle {\vec
  k}, -{\vec k} \mid V_{I=0}(E=2\mu) \mid {\vec k}^\prime, -{\vec k}^\prime
\rangle \; \frac {Z(k^\prime)} {2(\omega(k^\prime)-\mu)} \; \coth \frac
{\omega(k^\prime)-\mu}{2T_0^u}~ ~ ~,\label{ZEqn}
\end{eqnarray}
which is precisely the same condition as for $T_c^u$, Eq.~(\ref{thou1}).
The gap equation (\ref{gapeqn}) thus reduces to the $T$--matrix pole
condition at the particular energy $E=2\mu$.
In fermion systems, this is the well--known Thouless criterion
\cite{Thouless} for the onset of a phase transition to a pair
condensate. We note that the
Thouless criterion is only approximately valid if relativistic
corrections are included.

Several observations can be made. Firstly, one always obtains a pole
in the $T$--matrix 
if the temperature lies above $T_0^\ell(\mu)$.
Thus, at fixed $\mu$, no matter how weak the
interaction strength is (provided it is attractive in the neighborhood
of the $2\pi$ threshold), one always obtains a pole for 
sufficiently high temperatures (for fermions at a sufficiently
low temperature). In practice, $T_0^\ell$ (and $T_0^u$) will exceed sensible
values for pions as soon as $\mu$ drops below $\sim 130~{\rm MeV}$,
since they are increasing functions of $\mu$. Secondly, 
for a fixed interaction strength, the pole position
shifts downward with increasing temperature (for fermions: pole position
moves up with increasing temperature).
As a function of temperature, we therefore see a behavior for bosons
opposite to that for fermions.

The fact that increasing temperature reinforces the binding is
somewhat counterintuitive, but is an immediate consequence of the
coth--factor associated with bose statistics in Eq.~(\ref{A:Tmat2}).
Indeed, one realizes that the coth--factor increases with increasing
temperature and thus effectively enhances the two--body interaction.
We can therefore always find a bound state for arbitrarily small
attraction: it suffices to increase the temperature or, equivalently,
the density accordingly. This is opposite to the fermion case where
the corresponding tanh--factor suppresses the interaction with
increasing temperature. Therefore, in the fermion case, even at the
$T$--matrix level there exists a critical temperature where the Cooper
pole ceases to exist. For bosons, on the other hand, once one has
reached $T=T_0^\ell$, a bound state (here $E^{2\pi}_B < 2m_\pi$)
exists and the bound state energy simply continues to decrease as the
temperature increases. Of course, this becomes unphysical as soon as
the density of pairs becomes so strong that the bound states start to
obstruct each other, and finally dissolve at an upper critical
temperature (Mott effect). Precisely this non--linear effect is very
efficiently taken care of in the gap equation. In spite of the fact
that we still have a coth--factor in the gap equation, there is now a
crucial difference: the argument of the coth--factor is the
quasiparticle energy, Eq.~(\ref{Ek}), (over $T$) and thus, due to the
presence of $-\Delta^2(k)$ in $E(k)$, the origin of the coth is
shifted to the right with respect to the $T$--matrix case. Now, as $T$
increases, the only way to keep the equality of the gap equation is
for $\Delta(k)$ to decrease -- this pushes the origin of the coth back
to the left, counterbalancing its increase due to the increasing
temperature. Of course, this only works until $\Delta=0$, {\it i.e.},
until the temperature has reached $T_c^u$. This is precisely the
temperature $T_0^u$ for which the bound state in the $T$--matrix
reaches an energy $E_B^{2\pi}=2(m_\pi-\mu)$. We therefore see that in
spite of the fact that the bosons prefer high phase space density, the
formation of bound states ceases to exist beyond a critical
temperature -- just as for fermions.

Lastly, we return to the behavior of the pole for varying total
momentum $K$, and the threshold effect seen in Fig.~4. Since
$F_{\mu,T}(s,K)$ becomes complex above threshold, we show in Fig.~6
its magnitude for fixed $T=100~{\rm MeV}$ and $\mu=135~{\rm MeV}$, and
various values of $K$.  As expected, for increasing $K$ ({\it i.e.},
decreasing phase space occupation felt by the two pions in question)
the pole (zero of $F$) moves up in energy until it dissapears at some
critical momentum $100~{\rm MeV}/c < K_c < 250~{\rm MeV}/c$. For
$K>K_c$, the now non--zero minimum of the determinant function
continues to shift to higher energies, corresponding roughly to the
similar shift in the resonance peak in Fig.~4.


\section{Discussion and conclusions}

In the previous section, we investigated the effect of a thermal
medium on the pion dispersion relation at low momenta $k$.  In
particular, one finds a critical temperature $T_c^\ell$ at which the
pion dispersion relation is linear (phonon-like) in $k$ for small $k$.
This result is independent of the details of the interaction and
characteristic of any bose system (see Ref.~\cite{Evans2} for the case
of $^4$He). 

Such a change in the pion dispersion relation at low temperatures
would influence the pion spectra at low momentum.  For this to occur,
rather large medium phase space occupation numbers are required. In
particular, for a physically reasonable system with, say, $T <
200~{\rm MeV}$, this means that we require large chemical potentials.
In fact, dynamical calculations \cite{WB,BDSW} show that a buildup of
$\mu$ can indeed occur, provided that the scattering rate is
sufficiently large compared to expansion rate and the inelastic
collisions have ceased to be a factor.

To demonstrate the possible effect in a qualitative way, consider
the pion transverse momentum spectrum for longitudinally boost invariant
expansion
\begin{equation}\label{C22}
\frac{d N}{m_t d m_t dy} \;=\;
(\pi R^2 \tau) \frac{m_{\pi}}{(2 \pi)^2}
\sum^{\propto}_{n=1}\:
\exp({\frac{n \mu}{T}})\, K_1(\frac{n m_t}{T}) ~ ~,
\end{equation}
where $K_1$ is a Mc--Donald function, $m_t$ is the transverse mass,
and the normalization volume $\pi R^2 \tau$ is of the order $200$ to
$300~{\rm fm^3}$ \cite{Kataja}.  At mid-rapidity, the transverse mass
coincides with the full energy of the pion, and we follow
Ref.~\cite{Chanfray1} in replacing $m_t$ by the in--medium pion
dispersion relation $E(k)+\mu$ derived in the previous section. Of
course, as remarked in Ref.~\cite{Chanfray1}, the use of this
procedure is rather tenuous since the system is by definition still
far from freeze--out. In a dynamical calculation, hard collisions
would re--thermalize the system at ever decreasing temperatures.

In Fig.~7, the thermal transverse momentum spectrum for pions with
$\mu=135~{\rm MeV}$ and $T=100~{\rm MeV} > T_c^\ell(\mu)$ is shown
with (solid line) and without (dashes) the effect of the gap energy.
Essentially, the presence of a large chemical potential gives the
spectrum the appearance of one for small--mass particles, and the gap
energy, which causes $E(k) \sim k$ for long wavelengths, strengthens
this effect. We would like to mention here again that the use of our
force, Eq.~(\ref{0PV}), which respects chiral symmetry constraints
\cite{Drop}, considerably reduces the effect of binding with respect
to a purely phenomenological interaction, fitted to the phase shifts
(see, for example, Ref.~\cite{Drop}). This stems from the fact that
the expression (\ref{0PV}) becomes repulsive sufficiently below the
$2m_\pi$ threshold. This is not the case for commonly employed
phenomenological forces \cite{Drop}. As a consequence, the effect we
see in Fig.~7 is relatively weak, but one should remember that the
force (\ref{0PV}) is by no means a definitive expression.  It is well
known that in a many body system screening effects should be taken
into account. Whereas in a fermi system this tends to weaken the
force, it is likely that screening strengthens it in bose systems.  In
this sense our investigation can only be considered schematic. A
quantitative answer to the question of bound state formation in a hot
pion gas is certainly very difficult to give. Qualitatively, the
curves in Fig.~7 agree with the trend in the pion data at SPS
\cite{NA35} to be ``concave--up,'' but this is mainly an effect from
the finite value of the chemical potential \cite{WB,Kataja}. While the
gap changes the spectrum by a factor of $\sim 3$ at $m_t-m_\pi \sim
0$, this region is not part of detector acceptances.

In summary, we have shown that finite temperature induces real poles
in the $2\pi$ propagator below the $2m_\pi$ threshold, even for
situations where there is no $2\pi$ bound state in free space
\cite{CSW}. The situation is analogous to the Cooper pole of fermion
systems, and we therefore studied the corresponding bosonic ``gap''
equation. This equation has non--trivial solutions in a certain domain
of the $\mu$--$T$ plane.  Such a region always exists, even in the
limit of infinitesimally weak attraction. This is different from the
$T=0$ case discussed by Nozi{\`e}res and Saint James \cite{StJNoz},
where a nontrivial solution to the gap equation only exists when there
is a two boson bound state in free space. Our study has to be
considered preliminary. The final aim will be to obtain an equation of
state for a hot pion gas within a
Br\"uckner--Hartree--Fock--Bogoliubov approach. Also, the subtle
question of single boson versus pair condensation must be addressed
(see Ref.~\cite{StJNoz} and references therein). Furthermore, the fact
that we obtain two pion bound states in a pionic environment leads to
the speculation that higher clusters, such as four--pion bound states,
{\em etc.}, may also occur, and perhaps even be favored over pair
states. Such considerations, though interesting, are very difficult to
treat on a quantitative basis. However, substantial progress towards
the solution of four body equations has recently been made
\cite{PeterPriv}, and one may hope that investigations for this case
will be possible in the near future.

We are grateful to P.~Danielewicz for useful discussions.
This work was supported in part by the U.S. Department of Energy
under Grant No. DE-FG02-93ER40713.

\renewcommand{\theequation}{A.\arabic{equation}}
\setcounter{equation}{0}
\section{Appendix: Derivation of $T$--matrix and gap equation}

This appendix is devoted to a derivation of the gap equation for a
bosonic system governed by a field--theoretic Hamiltonian. The basic
problem one has to deal with is the formal introduction of a chemical
potential for bosons, since the total bosons number operator ({\it i.e.},
$\pi^++\pi^-+\pi^0$) does not commute with the Hamiltonian. Hence, if
we consider a pion gas at a typical temperature of $200~{\rm MeV}$,
it will correspond to zero chemical
potential. However, for a system lifetime on the order of tens of
fermi, the inelastic collision rate is negligible. Therefore,
provided the elastic collision rate is sufficiently large, a thermal
equilibrium with a finite chemical potential may well be reached.

Let us consider a pion system at temperature $T$.
Inspired by the
linear $\sigma$--model, with form factors fitted to the $\pi$--$\pi$
phase shifts, we take the Hamiltonian
\begin{equation}
H \;=\; H_0 \:+\: H_{\rm int}~ ~ ~,\label{A:H}
\end{equation}
where $H_0$ is the kinetic Hamiltonian for the $\pi$ and ``$\sigma$''
mesons
\begin{equation}
H_0 \;=\; \sum \,\omega_1 \, b_1^\dagger b_1 \:+\: \sum \,
\Omega_\alpha \, \sigma_\alpha^\dagger \sigma_\alpha~ ~ ~.\label{A:H0}
\end{equation}
The index ``1'' refers to the momentum and isospin of the pion, and
``$\alpha$'' to the momentum and identity of the heavy meson carrying
the interaction. The interaction Hamiltonian has the form
\begin{eqnarray}
H_{\rm int} &=& \frac 12 \, \sum\:
\bigg [ (\sigma_\alpha+\sigma_{-\alpha}^\dagger)\:
(b_1^\dagger b_2^\dagger+b_{-1}b_{-2}) \bigg ] \: \langle 12
\mid W \mid \alpha \rangle \nonumber \\
&+& \frac 14 \sum\: \bigg [b_1^\dagger b_2^\dagger b_3 b_4 \:+\: \frac 12
(b_{-1} b_{-2} b_3 b_4 + b_1^\dagger b_2^\dagger b_{-3}^\dagger
b_{-4}^\dagger)\bigg ] \: \langle 12 \mid V \mid 34 \rangle \label{A:Hint}
\end{eqnarray}
In the linear $\sigma$--model one has ($L^3$ is a normalization
volume)
\begin{eqnarray}
\langle 12 \mid W \mid \alpha \rangle &=& \bigg [ \frac {1} {
  2\omega_1 L^3\, 2\omega_2 L^3\, 2\Omega_\alpha L^3} \bigg ]^{1/2} \:
v(k^*_{12}) \; (2\pi)^3\delta({\vec k}_1+{\vec k}_2-{\vec k}_\alpha)
\;\frac {M_\sigma^2-m_\pi^2}{f_\pi} \; \delta_{12}
\label{A:W}\\
\langle 12 \mid V \mid 34 \rangle &=& \bigg [ \frac {1} {
  2\omega_1 L^3\, 2\omega_2 L^3\, 2\omega_3 L^3\,
2\omega_4 L^3} \bigg ]^{1/2} \: v(k^*_{12})\:v(k^*_{34})
\; (2\pi)^3\delta({\vec k}_1+{\vec k}_2-{\vec k}_3-{\vec k}_4)
\nonumber \\ &\times& \frac {M_\sigma^2-m_\pi^2}{f_\pi^2} \; \bigg [
\delta_{12}\,\delta_{34} \:+\: \delta_{13}\,\delta_{24} \, \frac
{2\omega_1\omega_3 -2{\vec k}_1\cdot{\vec k}_3 - m_\pi^2}{M_\sigma^2}
\:+\: \delta_{14}\,\delta_{23} \, \frac
{2\omega_1\omega_4 -2{\vec k}_1\cdot{\vec k}_4 -
  m_\pi^2}{M_\sigma^2}\bigg ]~ ~ ~.
\label{A:V}
\end{eqnarray}
The form factor taken at c.m. momentum $k^*$ of the pion pair is fitted
to the experimental phase shifts, and $M_\sigma$ is the
$\sigma$--mass. The static quartic interaction contains the
$\pi^2\pi^2$ interaction of the $\sigma$--model, and the $t$ and $u$
channel $\sigma$--exchange terms. We neglect the $t$ and $u$
dependence of the denominator (see Ref.~\cite{Drop}),
since their effect is extremely small. Further, terms like $\sigma
b^\dagger b$ and $b^\dagger b b b$ have been dropped, since they are
not essential for our purpose.

{\large {\it The Dyson equation for the pion propagator}}

We now derive the equation of motion for the pion propagator
\begin{equation}
G_{1{\bar 1}}(t,t^\prime) \;=\; \bigg \langle -i T\bigg
( b_1(t) b_1^\dagger(t^\prime)\bigg )\bigg \rangle \label{A:G11}
\end{equation}
where the $b_1$ are normal Heisenberg operators
$b_1(t)=\exp(iHt)b_1(0)\exp(-iHt)$. In principle, the extension to finite
temperature requires a matrix formulation (real time formulation) or
Matsubara Green's function. However, for simplicity we consider here the
normal zero temperature $G$, and replace it by a thermal propagator at
the end. We have checked that the final result is not modified.

After standard manipulation, using the Hamiltonian (\ref{A:H}), we obtain
\begin{equation}
\bigg (i \frac {\partial}{\partial t} - \omega_1 \bigg ) G_{1{\bar
    1}}(t,t^\prime) \;=\; \delta(t-t^\prime) \:+\: \int
dt^{\prime\prime} \: \Sigma_1(t,t^{\prime\prime})\: G_{1{\bar
    1}}(t^{\prime\prime}, t^\prime)~ ~ ~,\label{A:EoMG}
\end{equation}
with $\Sigma_1(t,t^\prime)=\Sigma_1^S(t,t^\prime)+
\Sigma_1^D(t,t^\prime)$. The static part of the mass operator is
\begin{equation}
\Sigma_1^S(t,t^\prime) \;=\; \sum_2\: \langle b_2^\dagger b_2 \rangle \:
\langle 12 \mid V \mid 12 \rangle \: \delta(t-t^\prime)~ ~
~,\label{A:MS}
\end{equation}
while the dynamical part is given by
\begin{equation}
\Sigma_1^D(t,t^\prime) \;=\; \bigg \langle -iT\bigg ( [H_{\rm int},b_1](t)\:
[H_{\rm int},b_1]^\dagger(t^\prime) \bigg ) \bigg \rangle~ ~ ~.\label{A:MD}
\end{equation}
Making a standard factorization approximation, we obtain
\begin{eqnarray}
\Sigma_1^D(t,t^\prime) &=& i\sum_2 G_{{\bar 2}2}(t,t^\prime)\nonumber \\
&\times& \bigg\langle -iT\bigg ( \bigg [\langle 12 \mid W\mid\alpha\rangle
\:(\sigma_\alpha+ \sigma_{-\alpha}^\dagger)(t) \:+\: \frac 12 \langle 12
\mid V \mid 34\rangle \:(b_3b_4+b_{-3}^\dagger b_{-4}^\dagger)(t)\bigg ],
\nonumber \\
&\mbox{}& \bigg[(\sigma_{\alpha^\prime}^\dagger+
\sigma_{-\alpha^\prime}^\dagger)(t^\prime)\:
\langle \alpha^\prime \mid W\mid 12\rangle \:+\: \frac 12
(b^\dagger_{3^\prime} b^\dagger_{4^\prime} + b_{-3^\prime}
b_{-4^\prime})(t^\prime)\:\langle 3^\prime 4^\prime \mid V \mid 12
\rangle \bigg ] \bigg )\bigg\rangle~ ~ ~,\label{A:MD2}
\end{eqnarray}
with $G_{{\bar 2}2}(t,t^\prime) = \langle -i T
(b^\dagger_2(t),b_2(t^\prime)) \rangle$, and there is an implicit
summation over repeated indices.

{\large {\it Extraction of the condensates}}

In the above expression the operator $b_3^\dagger b_4^\dagger$
connects states with $N$ particles to states with $N+2$
particles. Among these states those with excitation energy $2\mu$ play
a prominent role (Cooper poles). To separate the influence of these
states, we split the fluctuating part of the operator from the
condensate
\begin{equation}
b_3^\dagger b_4^\dagger(t) \;=\; \langle b_3^\dagger b_4^\dagger(t)
\rangle  \:+\: :b_3^\dagger b_4^\dagger(t):~ ~ ~ ~. \label{A:Split}
\end{equation}
The time evolution is
\begin{equation}
 \langle b_3^\dagger b_4^\dagger(t)\rangle \;=\;  \langle b_3^\dagger
b_{-3}^\dagger\rangle \; {\rm e}^{i2\mu t} \;
\delta_{3,-4}~ ~ ~,\label{A:timeEv}
\end{equation}
where  $\langle b_3^\dagger b_{-3}^\dagger \rangle$ is the usual time
independent pion density. Similarly, we obtain
\begin{equation}
b_3 b_4(t) \;=\; \langle b_3 b_{-3} \rangle  \; {\rm e}^{-i2\mu t} \;
\delta_{3,-4} \:+\: :b_3 b_4(t):~ ~ ~ ~. \label{A:Split2}
\end{equation}
We now extract the condensate part of the $\sigma$--field operator from
the fluctuating part:
\begin{equation}
\sigma_\alpha(t) \;=\; \langle \sigma_\alpha(t) \rangle \:+\:
s_\alpha(t)~ ~ ~. \label{A:SigCon}
\end{equation}
The equation of motion gives
\begin{equation}
i \frac{\partial}{\partial t} \langle \sigma_\alpha(t) \rangle \;=\;
\Omega_\alpha \langle \sigma_\alpha(t) \rangle \:+\: \frac 12
\langle (b_1b_2 + b_{-1}^\dagger b_{-2}^\dagger \rangle \:
\langle\alpha\mid W\mid 12\rangle~ ~ ~.\label{A:SigEoM}
\end{equation}
We look for a solution of the form
\begin{equation}
\langle \sigma_\alpha(t) \rangle \;=\; ( A\, {\rm e}^{-i2\mu t} \:+\:
B\,{\rm e}^{i2\mu t} )\: \delta_{\alpha 0}~ ~ ~.\label{A:Sol}
\end{equation}
$A$ and $B$ are straightforwardly obtained from the equation of motion:
\begin{equation}
\langle \sigma_0(t) \rangle \;=\; -\frac 12\: \frac {\langle b_1 b_{-1}
  \rangle \langle 0\mid W\mid 1-1\rangle}{M_\sigma-2\mu} \:
{\rm e}^{-i2\mu t}
\;-\;\frac 12\: \frac {\langle b_1^\dagger b_{-1}^\dagger
  \rangle \langle 0\mid W\mid 1-1\rangle}{M_\sigma+2\mu} \:
{\rm e}^{i2\mu t}~ ~ ~.\label{A:Sol2}
\end{equation}
In the expression of the dynamical mass operator one can extract a
Cooper pole part, where only the condensates occur. The remaining part
involves only the fluctuating pieces. Grouping the latter with the
static mass operator, we can write
\begin{equation}
\Sigma_1(t,t^\prime) \;=\; \Sigma_{1C}(t,t^\prime) \:+\:
\Sigma_{1H}(t,t^\prime)~ ~ ~,\label{A:MC-H}
\end{equation}
where $\Sigma_{1H}(t,t^\prime)$ is the normal Hartree mass operator which
depends on the full in--medium $T$--matrix:
\begin{equation}
\Sigma_{1H}(t,t^\prime) \;=\; i \sum_2 \: G_{{\bar 2}2}(t,t^\prime) \:
\langle 12 \mid T(t,t^\prime) \mid 12 \rangle~ ~ ~.\label{A:MtoT}
\end{equation}
with
\begin{eqnarray}
\langle 12 \mid T(t,t^\prime) \mid 34 \rangle
 &=& \langle 12 \mid V \mid 34 \rangle \: \delta(t-t^\prime)\nonumber\\
&+&  \bigg\langle -iT\bigg ( \bigg [\langle 12 \mid W\mid\alpha\rangle
\:(s_\alpha+ s_{-\alpha}^\dagger)(t) \:+\: \frac 12 \langle 12
\mid V \mid 56\rangle \: (:b_5b_6+b_{-5}^\dagger b_{-6}^\dagger :)(t)\bigg ],
\nonumber \\
&\mbox{}& \bigg[(s_{\alpha^\prime}^\dagger+
s_{-\alpha^\prime}^\dagger)(t^\prime)\:
\langle \alpha^\prime \mid W\mid 34\rangle \:+\: \frac 12
(:b^\dagger_{5^\prime} b^\dagger_{6^\prime} + b_{-5^\prime}
b_{-6^\prime}:)(t^\prime)\:\langle 5^\prime 6^\prime \mid V \mid 34
\rangle \bigg ] \bigg )\bigg\rangle~ ~ ~\label{A:TMat1}
\end{eqnarray}
Using the Dyson equation for the $b$ and $s$ operators, it is a purely 
technical matter to show that this scattering amplitude satisfies a
Lippmann--Schwinger equation. In energy space, and in the $I=0$ 
channel, it reads:
\begin{equation}
\langle 12 \mid T_{I=0}(E) \mid 34 \rangle \;=\; \langle 12 \mid
V_{I=0}(E) \mid 34 \rangle \;+\; \frac 12\, \langle 12 \mid
V_{I=0}(E) \mid 56 \rangle \: G_{2\pi}^{56}(E) \: \langle 56 \mid
T_{I=0}(E) \mid 34 \rangle~ ~ ~,\label{A:LSE}
\end{equation}
where $G_{2\pi}(E)$ is the in--medium $2\pi$ propagator
\begin{equation}
G_{2\pi}^{56}(E) \;=\; \bigg [ \frac {1} {E - (\omega_5+\omega_6)
  +i\eta} \;-\;  \frac {1} {E + (\omega_5+\omega_6) + i\eta}
\bigg ] \; (1\:+\:f_5\:+\:f_6)~ ~ ~,\label{A:2piProp}
\end{equation}
with thermal occupation numbers $f(k)=[ \exp (\omega(k)-\mu)/T - 1 ]^{-1}$.
As mentioned above, we have checked that using the correct matrix
form of the two pion propagators instead of (\ref{A:2piProp}) yields
the same final result. In Eq.~(\ref{A:LSE}), $V_{I=0}(E)$ is the
effective $\pi$--$\pi$ potential in the $I=0$ channel which 
incorporates all the tree level diagrams. For total incoming momentum
${\vec K} = {\vec k}_1+{\vec k}_2 = {\vec k}_3+{\vec k}_4$, it reads
\begin{equation}
\langle {\vec k}_1, {\vec k}_2 \mid V_{I=0}(E) \mid {\vec k}_3, {\vec
  k}_4 \rangle \;=\; 
\bigg ( \frac {1}{2\omega_12\omega_22\omega_32\omega_4}\bigg )^{1/2}
\: \langle {\vec k}_1, {\vec k}_2 \mid {\cal M}_B(E) \mid {\vec k}_3, {\vec
  k}_4 \rangle~ ~ ~,\label{InvMB}
\end{equation}
where the bare invariant interaction ${\cal M}_B$ is
\begin{equation}
\langle {\vec k}_1, {\vec k}_2 \mid {\cal M}_B(E) \mid {\vec k}_3, {\vec
  k}_4 \rangle \; \equiv \;
\langle k_{12}^* \mid {\cal M}_B(s) \mid k_{34}^*\rangle
\;=\; \sum_{i=1}^{2}\: \lambda_i(s)\: v_i(k_{12}^*)\,v_i(k_{34}^*)
~ ~ ~,\label{A:0PV}
\end{equation}
with
\begin{eqnarray}
v_1(k) &=& v(k) \equiv [1+(k/8m_\pi)^2]^{-1}~ ~, ~ ~ ~ ~ ~
v_2(k) \;=\; \frac{\omega(k)}{m_\pi} \:v(k)~ ~, \label{A:FormFactor}\\
\lambda_1(s) &=& \frac {M_\sigma^2-m_\pi^2}{f_\pi^2} 
\; \bigg [ 3 \: \frac {s-m_\pi^2}{s-M_\sigma^2} \:-\: \frac
{2m_\pi^2}{M_\sigma^2} \bigg ]~ ~, ~ ~ ~ ~ ~
\lambda_2 \;=\; \frac {M_\sigma^2-m_\pi^2}{f_\pi^2}
\; \frac{4m_\pi^2}{M_\sigma^2}  ~ ~ ~. \label{A:Vab}
\end{eqnarray}
In these equations 
$s=E^2-{\vec K}^2$ is the square of the total c.m. energy, and the
$k_{ij}^*$ are the magnitudes of the relative 3--momenta in the c.m. 
frame
\begin{eqnarray}
\omega_{ij}^{*\,2} &=& m_\pi^2 \:+\: {\vec k}_{ij}^{*\,2} \;=\; \frac 14
\bigg [ (\omega_i + \omega_j)^2 \:-\: {\vec K}^2 \bigg ]~ ~ ~, ~
i,j=1,2~{\rm or}~3,4. \nonumber
\end{eqnarray}
The form factor $v(k)$, Eq.~(\ref{A:FormFactor}), 
and $\sigma$--mass $M_\sigma = 1~{\rm GeV}$ 
have been fitted to the experimental phase shifts.

The Lippmann--Schwinger equation for the invariant $T$--matrix, 
${\cal M}_{I=0}$, may finally be rewritten in a form suitable for 
practical purposes:
\begin{eqnarray}
\langle k_{12}^* \mid {\cal M}_{I=0}(E,K) \mid k_{34}^* \rangle &=& 
\langle k_{12}^*\mid {\cal M}_B(s)\mid k_{34}^*\rangle\nonumber\\
&+& \frac 12 \,\int \! \frac {d^3k_{56}^*}{(2\pi)^3} \;
\langle k_{12}^* \mid {\cal M}_B(s)  \mid k_{56}^*\rangle \;
\frac {1}{\omega_{56}^*} \: \frac {\langle 1+f_++f_-\rangle}
{s-4\omega_{56}^{*\,2}+i\eta} \; 
\langle k_{56}^* \mid {\cal M}_{I=0}(E,K)
\mid k_{34}^* \rangle~ ~ ~.\label{A:Tmat2}
\end{eqnarray}
In the special case of a single fireball of temperature $T$ and
chemical potential $\mu$, the angle average factor is given by
\begin{equation}
\langle 1+f_++f_-\rangle \;=\; \frac {T}{\gamma \beta k^*_{56}} \: \ln
\frac {\sinh [\{\gamma(\omega^*_{56}+\beta k^*_{56})-\mu\}/2T]}
      {\sinh [\{\gamma(\omega^*_{56}-\beta k^*_{56})-\mu\}/2T]}~ ~ ~,
\label{A:Angle} 
\end{equation}
where $\beta$ and $\gamma$ are the velocity and gamma--factor of the
pair with respect to the bath. This factor reduces to $\coth
(\omega_{56}^*-\mu)/2T$ for vanishing incoming total momentum ${\vec K}$.

Eq.~(\ref{A:Tmat2}) is solved by a separable ansatz
\begin{equation}
\langle k_{12}^* \mid {\cal M}_{I=0}(E,K) \mid k_{34}^* \rangle \;=\;
\sum_{i=1}^2 \: \lambda_i(s) \: v_i(k_{12}^*)\,\tau_i(k^*_{34};s, {\vec
  K})~ ~ ~, \label{A:MSol}
\end{equation}
where the functions $\tau_i$ obey the coupled set of equations
\begin{equation}
\sum_{j=1}^2 \; \bigg [ \delta_{ij}\:-\; \lambda_j(s)\,g_{ij}(s,{\vec
  K}) \bigg ] \:
\tau_j(k; s,K) \;=\; v_i(k)~ ~,~ ~ ~ ~ ~i=1,2 \label{A:MME}
\end{equation}
with
\begin{equation}
g_{ij}(s,K) \;\equiv\; \frac 12 \: \int \frac{d^3k}{(2\pi)^3} \;
v_i(k) \: \frac {1}{\omega(k)} \, \frac {\langle 1+f_++f_- \rangle}
{s-4\omega_k^2 +i\eta} \: v_j(k)~ ~ ~.\label{A:Deffij}
\end{equation} 

{\large {\it The gap equation}}

To obtain the Cooper piece of the mass operator we must simply
replace $\sigma$ and $b b$ by $\langle \sigma \rangle$ and
$\langle b b\rangle$. According to the previous result, we find after
some straightforward algebra, and noting that the index $2$ is
necessarily $-1$,
\begin{equation}
\Sigma_{1C}(t,t^\prime) \;=\; -i\,G_{-{\bar 1},-1}(t,t^\prime) \: F_1^2 \:
\bigg ( {\rm e}^{-i2\mu(t-t^\prime)} \:+\: {\rm e}^{i2\mu(t-t^\prime)}
\bigg  )~ ~ ~.\label{A:M1C}
\end{equation}
The important point is that $F_1$ involves the $I=0,\ell=0$ energy 
dependent $\pi$--$\pi$ potential at $E = 2\mu$:
\begin{equation}
F_1 \;=\; -\frac 12 \: \int \! \frac {d^3k_2}{(2\pi)^3} \: \langle {\vec
  k}_1, -{\vec k}_1 \mid V_{I=0}(E=2\mu) \mid {\vec k}_2, -{\vec k}_2
\rangle \; \langle b_2 b_{-2} \rangle~ ~ ~.\label{A:F1}
\end{equation}
In energy space, $\Sigma_{1C}(\omega)$ is
\begin{eqnarray}
\Sigma_{1C}(\omega) &=& F_1^2 \:
\int \! d\tau \: {\rm e}^{i\omega\tau} \: \{
\theta(\tau)\, \langle b_1^\dagger(\tau), b_1(0) \rangle \:+\:
\theta(-\tau) \, \langle b_1(0), b_1^\dagger(\tau) \rangle \}
\nonumber \\
&\mbox{     }& \times ({\rm e}^{-i2\mu\tau} \:+\: {\rm e}^{i2\mu\tau}
)~ ~ ~. \label{A:M1CE}
\end{eqnarray}
Taking for $b^\dagger_1(\tau)$ the bare time evolution
$b_1^\dagger(\tau) = b_1 {\rm e}^{i\omega\tau}$, and keeping only the
real part, we finally find
\begin{equation}
\Sigma_{1C}(\omega) \;=\; -F_1^2 \: \bigg ( \frac {1}{\omega+\omega_1-2\mu}
\:+\: \frac {1}{\omega+\omega_1+2\mu} \bigg )~ ~ ~.\label{A:M1CEF}
\end{equation}
The first term is the usual non--relativistic result, and the second
one corresponds to a relativistic correction.

Reinserting the result (\ref{A:M1CEF}) into the Dyson equation for the
pion propagator, and ignoring the Hartree correction, we find that the
pole of the pion propagator is the solution of
\begin{equation}
(\omega-\mu)^2 \;=\; (\omega_1-\mu)^2 \:-\: F_1^2 \:\bigg [ 1 \: +\:
\frac {(\omega-\mu)+(\omega_1-\mu)}{(\omega-\mu)+(\omega_1-\mu)+4\mu}
\bigg ]~ ~ ~,\label{A:Soln}
\end{equation}
The second term in the square brackets represents a relativistic
correction to the standard dispersion relation, since for typical
non--relativistic situation one has
\begin{equation}\nonumber
\mu\; {\buildrel < \over \sim} \; m_\pi \mbox{,~ ~ ~ ~} \omega-m_\pi
\ll m_\pi \mbox{, ~ ~and~ ~ ~} \omega_1 - m_\pi \ll m_\pi~ ~ ~.
\end{equation}

Calling
\begin{equation}
\Delta_1^2 \;=\; F_1^2\: \bigg  [ 1 \: +\:
\frac {E_1+(\omega_1-\mu)}{E_1+(\omega_1-\mu)+4\mu}
\bigg ]~ ~ ~,\label{A:DDef}
\end{equation}
the quartic equation can be approximated by a quadratic one in terms
of $\omega-\mu$:
\begin{equation}
E_1^2 \;=\; (\omega-\mu)^2 \:-\: \Delta_1^2~ ~ ~,\label{A:E1Def}
\end{equation}
with a gap equation following from Eq.~(\ref{A:F1}) and (\ref{A:M1CEF})
\begin{equation}
\Delta_1 \;=\; -\frac 12 \:\int \! \frac {d^3k_2}{(2\pi)^3} \: \langle {\vec
  k}_1, -{\vec k}_1 \mid V_{I=0}(E=2\mu) \mid {\vec k}_2, -{\vec k}_2
\rangle \; \frac {\Delta_2} {2E_2} \; \coth \frac {E_2}{2T} \; \bigg [
1 \:+\: \frac {E_1+(\omega_1-\mu)}{E_1+(\omega_1-\mu)+4\mu} \bigg ]^{1/2}~
~ ~,\label{A:GapEqn}
\end{equation}
which is the standard gap equation with a relativistic correction.
The presence of the factor $1/2$ is somewhat unconventional, but is
simply related to the fact that the matrix element of the interaction 
incorporates the exchange term. Note that the factor $1/4$ in front of
the quartic term of the interaction Hamiltonian Eq.~(\ref{A:Hint}) has
the same origin.

To calculate the occupation number, we note, using the explicit
form of $G_{1{\bar 1}}$, that the bare vacuum is the vacuum of
quasi--particle operators $B_1$, such that
\begin{equation}
b_1 \;=\; \bigg [ \frac {\omega_1-\mu}{2E_1} \:+\: \frac 12 \bigg
]^{1/2}\: B_1 \;+\; \bigg [  \frac {\omega_1-\mu}{2E_1} \:-\: \frac 12
\bigg ]^{1/2}\: B_{-1}^\dagger~ ~ ~.\label{A:btoB}
\end{equation}
Using $\langle B^\dagger B \rangle = [\exp(E_1/T) - 1 ]^{-1}$, it 
follows that
\begin{eqnarray}
\langle b_1^\dagger b_1 \rangle &=& \frac {\omega_1-\mu}{2E_1}\:
\coth \frac {E_1}{2T} \;-\; \frac 12 \nonumber \\
\langle b_1 b_{-1} \rangle &=& \frac{\Delta_1}{2E_1} \; \coth
\frac {E_1}{2T}~ ~ ~.\label{A:Dens}
\end{eqnarray}

{\large {\it The Thouless Criterion}}

We may also obtain the condition for having a pole in the $T$--matrix
at $E=2\mu$. For $E$ near $2\mu$, we write
\begin{equation}
\langle {\vec k}_1, -{\vec k}_1 \mid T_{I=0}(E) \mid {\vec k}_2, -{\vec
  k}_2 \rangle \;\equiv \; Z_1\: \frac {1}{E-2\mu} \: Z_2~ ~
~.\label{A:TnearPole}
\end{equation}
Multiplying (\ref{A:Tmat2}) by $E-2\mu$ and taking the limit $E \to 2 \mu$,
one obtains an equation for $Z_1$
\begin{eqnarray}
Z_1 &=& -\frac 12 \:\int \! \frac {d^3k_2}{(2\pi)^3} \: \langle {\vec
  k}_1, -{\vec k}_1 \mid V_{I=0}(E=2\mu) \mid {\vec k}_2, -{\vec k}_2
\rangle \; \frac {Z_2} {2(\omega_2-\mu)} \; \coth \frac
{\omega_2-\mu}{2T} \nonumber \\ &\mbox{}&~ ~ ~ ~ ~ ~ ~ ~ ~ ~ ~ ~ ~ ~ ~
~ ~ ~ \times \bigg [
1 \:+\: \frac {E_2+(\omega_2-\mu)}{E_2+(\omega_2-\mu)+4\mu} \bigg ]~
~ ~,\label{A:ZEqn}
\end{eqnarray}
In the non--relativistic limit, {\it i.e.}, neglecting the last term
in the square brackets, this equation coincides with the linearized
form of the (non--relativistic) gap equation. This is just the
Thouless criterion: the gap equation begins to exhibit non--trivial
solutions at the point where the $T$--matrix has a pole at zero energy,
$\langle H-\mu N\rangle = E -2\mu =0$. We see that the
Thouless criterion is only approximately valid if relativistic
corrections are included.

\newpage



\centerline{\large {\bf FIGURE CAPTIONS}}

Fig.~1 The critical temperatures $T_c^u$ (solid line)
and $T_c^\ell$ (dashes).

Fig.~2 The gap strengths $\delta_1$ (solid) and $-\delta_2$ (dashes)
vs temperature $T$, for $\mu=135~{\rm MeV}$. See text.

Fig.~3 The quasiparticle dispersion relation at fixed $\mu=135~{\rm
  MeV}$, for three temperatures: $T_c^\ell$ (dashes), $T=115~{\rm
  MeV}$ (dot--dash), and $T_c^u$ (solid).

Fig.~4 The square of the on--shell invariant $T$--matrix 
for $I=0,\ell=0$ in free space (solid line). Also shown are the 
results in a thermal bath with $T=100~{\rm MeV}$, $\mu=135~{\rm MeV}$ 
for total momentum $K=0$ (short dashes), $200~{\rm MeV}/c$ (long dashes),
$1~{\rm GeV}/c$ (dot--dashes), and $3~{\rm GeV}/c$ (dotted).

Fig.~5 The $T$--matrix pole function $F_{\mu,T}(E)$ at various
temperatures for $\mu=135~{\rm MeV}$ and $K=0$. 
Here, $T_0^u=T_c^u\approx 123~{\rm MeV}$,
$T^\ell_c\approx 77~{\rm MeV}$, and $T_0^\ell \approx 47~{\rm MeV}$.

Fig.~6  The magnitude of $F_{\mu,T}(s,K)$, Eq.~(\ref{DetFunc}), for
a thermal medium with $T=100~{\rm MeV}$ and $\mu = 135~{\rm MeV}$.
The solid, short
dashed, long dashed, and dot--dashed lines correspond to $K=0$,
$100~{\rm MeV}/c$, $250~{\rm MeV}/c$, and $500~{\rm MeV}/c$, respectively. 

Fig.~7 The thermal pion transverse mass spectrum at midrapidity, 
for $T=100~{\rm MeV}$ and $\mu=135~{\rm MeV}$, without the 
effect of a gap (dashes) and including the effect of a gap (solid line).


\end{document}